\documentclass{article}
\usepackage[utf8]{inputenc}
\pdfoutput=1
\usepackage{amsmath}
\usepackage{mathrsfs}
\usepackage{mathtools}
\usepackage{amsthm}
\usepackage{semantic}
\usepackage{tikz-cd}
\usepackage{graphicx}
\usepackage{subfigure}
\usepackage{amsfonts}
\usepackage{amssymb}
\usepackage[backend=biber,style=authoryear,sorting=ynt]{biblatex}
\graphicspath{ {images/} }
\let\code=\texttt
\title{Simulating human interactions in supermarkets to measure the risk of COVID-19 contagion at scale}
\author{Serge Plata \and Sumanas Sarma \and Melvin Lancelot \and Kristine Bagrova \and David Romano-Critchley}
\date{June 2020}

\begin{document}

\maketitle

\section{Abstract}

Taking the context of simulating a retail environment using agent based modelling, a theoretical model is presented that describes the probability distribution of customer ``collisions" using a novel space transformation to the Torus $Tor^2$. A method for generating the distribution of customer paths based on historical basket data is developed. Finally a calculation of the number of simulations required for statistical significance is developed. An implementation of this modelling approach to run simulations on multiple store geometries at industrial scale is being developed with current progress detailed in the technical appendix. 

\section{Introduction}

This paper came as a result of the work developed for the London Royal Society's RAMP initiative to tackle the COVID-19 pandemic. 

Our main approach was a practical one: i) To simulate the risk of spreading the virus amongst the population when shopping at supermarkets and ii) Help shape policy on how to operate these businesses during the pandemic and lockdown exit-period.

This paper is focused on the mathematical foundations and premises to apply them on the computational solution i.e. the actual simulation engine.

The mathematical work is divided in 4 parts which solve in a practical way  fundamental problems for modelling in retail: i) The probability distribution of collisions, which came out of the discussion with the group and applying basic concepts of modelling, differential equations and statistics. We moved away from \cite{chen2012} as they assume a shortest path type behaviour which rarely happens in shopping trips ii) the dynamical system in a store and how to model it in order to get some useful metrics; for this we took advantage of the homeomorphisms on the Torus $Tor^2$, which allowed us to map the 2-dimensional space into the torus to simplify all calculations, which in the normal $\mathbb{R}^2$ would have been more complicated iii) the calculation of trajectories based on consumer baskets, which again came as a discussion with the group and involved the vectorization of customers, and an unsupervised algorithm to group them; and iv) the minimum sample size to achieve statistical significance on the simulation, which also came from a basic discussion with the group and applying foundational statistical knowledge, normally disregarded when implementing simulations and practical trials.

In this model we take a different approach to zonal graph representation of store geometry proposed in \cite{ying2020} which extends a mobility and congestion model \cite{ying2019}. The use of historical basket data to define customer paths is similar in both approaches.

\section{Mathematical basis and assumptions}

The model was developed using a simulation tool and feeding it with mathematical assumptions and models. It is important to note that the work was done out of discussions with the group and the amount of mathematical knowledge was varied and covered concepts of probability, dynamical systems, differential equations, and analytical geometry. We prioritized an approach that would allow these calculations to be performed at scale. All this comes from the classic mathematics theory and concepts and best computational practices.

\subsection{Probability}
\label{sec:probability}

One of the main objectives of this work was to assess the risk of spreading the virus. The risk in this case is measured as a probability. This is aligned with \cite{10.1287/mksc.1080.0400} as his rationale also considers events happening in time. Given our commercial experience in the field, it was clear for us that the events that lead to the spread of the virus were a function of the number of ``collisions" $x$ in a period of time $t$. A ``collision" happens when two people are within 2 metres distance from each other.  Hence the probability function that could model this should have 2 parameters: $x$ and $t$, and the function should be of the form $f(x, t)$. In general, this function should measure the rate at which events happen. 

The main goal of this task was to find the probability of $x$ collisions occurring in a time-interval of length $t$. 

The approach was to measure the number of collisions $x$ in an extended time-interval $t + \Delta t$. Using the notation above, the goal was to find the probability $f(x, t+ \Delta t)$. It was assumed that this probability is proportional to the length of time exposed. In other words it is more probable to get the virus if someone is ``exposed" for 4 hours than for 5 minutes. So, $f(x, t+ \Delta t) = \lambda \Delta t + \mu \Delta t$, where $\lambda \Delta t$ is the probability of one collision occurring in the time interval $\Delta t$ and $\mu \Delta t$ is the probability of observing more than one collision in $\Delta t$ and $\lambda$ and $\mu$ are the constant of proportionality.

One immediate corollary from this assumption is that the probability is constant over time. To go forward, another very important assumption was introduced: the uniqueness of events in time; in other words, only one collision can occur at any one moment. \par

This assumption leads immediately to the fact that $\mu \Delta t = 0$ (and simplifies our calculations meaningfully) as there are no simultaneous collisions, therefore the probability of the occurrence of \underline{one} collision is $f(1, t + \Delta t) = \lambda \Delta t$.

Further important assumptions are:

1. Independence: The probability of occurrence in one time interval is constant and has no relation with the probability of occurrences in other time intervals. \par
2. Uniform observation period: The observation period of all units is the same length of time.

Applying standard mathematical theory, and in order to assess the rate of spread we need to count the collisions $x$ that can occur in $t + \Delta t$ the first step is to record the number of collisions until time $t$ and then the number of collisions from $t$ to $t+ \Delta t$. Two things can happen: $x$ collisions occur between time $t=0$ and $t$ or the interval $[0, t]$, with none occurring in the interval $(t, t + \Delta t]$ or $x - 1$ collisions occur in $[0, t]$ and only one in $[t, t + \Delta t]$. 
Since we know the probability of observing one and only one collision in $[t,t + \Delta t]$ is  $\lambda \Delta t$, then the probability of $x$ collisions happening in the first interval $[0, t]$ is:
\begin{equation}
f(x, t) \cup f(0, t + \Delta t) = f(x, t)(1 - f(1, t + \Delta t)) = f(x, t)(1 - \lambda \Delta t) 
\end{equation}

and the probability of all the collisions is: \par
\begin{equation}
  f(x - 1, t) \cup f(1, t + \Delta t) = f(x - 1, t)f(1, t + \Delta t) = f(x - 1, t)\lambda \Delta t  
\end{equation}

Since the probability of collisions in this time interval is independent of collisions occurring in other intervals, we can write the probability of $n$ collisions in $[t, T+ \Delta t]$ as:

\begin{equation}
f(x, t + \Delta t) = f(x, t)(1 - \lambda \Delta t) + f(x - 1, t)\lambda \Delta t 
\end{equation}

which can be re-written as:

\begin{equation}
\frac{f(x, t + \Delta t) - f(x, t)}{ \Delta t} = - \lambda f(x, t)  +  \lambda f(x - 1, t)
\end{equation}

Taking the limit when $\Delta t \to 0$ we obtain a differential:

\begin{equation}
\lim\limits_{\Delta t \rightarrow 0} \frac{f(x, t + \Delta t) - f(x, t)}{\Delta t} = - \lambda f(x, t)  + \lambda f(x - 1, t) = \frac{df(x,t)}{dt}
\end{equation}

The above differential equation should be solved for all possible values of $x \in \mathbb{Z}$.

Thus, doing it in a discrete way,  and establishing our first boundary condition: $f(0, 0) = 1$ this means that the probability of collision at time 0 is 0, in other words ``no collisions occur at the start of the process". An equivalent proposition is that $f(x,0)=0$ for any $x \in \mathbb{Z}^{+}$.

Now, for the calculation of the probability of no collisions, i.e. for $x=0$ or $f(0, t)$ we just need to substitute $x = 0$ in equation 7:

\begin{equation}
\lim\limits_{\Delta t \rightarrow 0} \frac{f(0, t + \Delta t) - f(0, t)}{\Delta t} = - \lambda f(0, t)  + \lambda f(0 - 1, t) = \frac{df(0,t)}{dt}
\end{equation}

Since $f(-1,t)=0$ as per our initial condition, in other words only $x>0$ are allowed, we have:

\begin{equation}
\frac{df(0,t)}{dt} = - \lambda f(0, t)  + 0
\end{equation}

and rearranging:

\begin{equation}
\frac{df(0,t)}{f(0,t)} = - \lambda dt
\end{equation}

then integrating on both sides:

\begin{equation}
\int \frac{df(0,t)}{f(0,t)} = \int - \lambda dt
\end{equation}

\begin{equation}
ln [f(0, t)] = - \lambda t + C_0
\end{equation}

since the initial condition for the differential equation is $C_0=0$, we have that

\begin{equation}
f(0, t) = e^{- \lambda t}
\end{equation}

Following the same method, the calculation of the case for $x=1$, or in other words the probability of 1 collision in time $t$ or $f(1, t)$  comes by substituting again in equation 7:

\begin{equation}
\frac{df(1,t)}{dt} = - \lambda f(1, t)  + \lambda f(1 - 1, t)
\end{equation}

\begin{equation}
\frac{df(1,t)}{dt} = - \lambda f(1, t)  + \lambda f(0, t)
\end{equation}

Substituting $f(0, t)$ from equation 13:

\begin{equation}
\frac{df(1,t)}{dt} = - \lambda f(1, t)  + \lambda e^{- \lambda t}
\end{equation}

\begin{equation}
\frac{df(1,t)}{dt} + \lambda f(1, t)  = \lambda e^{- \lambda t}
\end{equation}

From the theory of ordinary differential equations, we multiply by the integrating factor: $\lambda e^{- \lambda t}$

\begin{equation}
\frac{ e^{ \lambda t} df(1,t)}{dt} + e^{ \lambda t} \lambda f(1, t)  = e^{ \lambda t} \lambda e^{- \lambda t}
\end{equation}

And integrating with respect to $t$:

\begin{equation}
\int \frac{ e^{ \lambda t} df(1,t)}{dt} + e^{ \lambda t} \lambda f(1, t)  = \int  \lambda
\end{equation}

\begin{equation}
f(1,t)  e^{ \lambda t} = \lambda t + C_1
\end{equation}

Since the next boundary value is when $f(1, 0)$ we have that $C_1 = 0$ and the result becomes:

\begin{equation}
f(1,t)  e^{ \lambda t} = \lambda t
\end{equation}

and finally:

\begin{equation}
f(1,t) = \lambda t e^{- \lambda t}
\end{equation}

In the same way the calculation of the following values for $x=2,3,4,...$ can follow:

\begin{equation}
f(2,t) =  \frac{e^{- \lambda t} \lambda^2 t^2}{2}
\end{equation}

\begin{equation}
f(3,t) =  \frac{e^{- \lambda t} \lambda^3 t^3}{6}
\end{equation}

And

\begin{equation}
f(4,t) =  \frac{e^{- \lambda t} \lambda^4 t^4}{24}
\end{equation}

and in its more general form, when $x=n$:

\begin{equation}
f(n,t) =  \frac{e^{- \lambda t} \lambda^n t^n}{n!}
\end{equation}

Which is a Poisson probability density function with mean $\lambda$, however in our equation, the mean depends on the time spent in the store. In other words we assume a Poisson process with mean that is directly proportional to the time spent in the store.

\subsection{Dynamics in a store}

To model the dynamics of customer behaviour in a store, it was decided to use some of the theory of chaotic dynamical systems and apply some classic metrics of this theory like the rotation number and recurrence concepts. This idea moves away from ideas like the \cite{ying2019} where the space is broken into pieces, however in spirit we have the same idea of a customer journey as a sequence but we added the time variable which is crucial to measure the simultaneity and therefore the risk of spread of shoppers in a store. 

The main idea is to ``see" the store from above and map it to a subset of $\mathbb{R}^2$ to then analyze the orbits of this system on that subset.

Then, in order to simplify the analysis of the possible trajectories of people on a shopping trip, we used the theory of homeomorphisms on the torus, firstly because the torus can be seen as the set of all equivalence classes of the points in $\mathbb{R}^2$ and secondly because different trajectories in different aisles in a store can be comprised into one.

The torus seen in this way can give us the information needed about the dynamic of the system which would be more laborious to do it on the plane.

The torus can be define as $S^1 \times S^1$, where $S^1$ is the unit circle and where all the properties of $S^1$ hold, thus each coordinate of the torus is represented in the form $( \theta_1 + 2 \pi r, \theta_2 + 2 \pi s  )$ with $r$ and $s \in \mathbb{Z}$. The mapping functions from $\mathbb{R}$ to the Torus, in this case can be seen in the following commutative diagram:
\begin{center}
\begin{tikzcd}
\mathbb{R}^2 \arrow{d}{(e^{2 \pi i x}, e^{2 \pi i y})} \arrow{r}{f=(f_1,f_2)}
& \mathbb{R}^2 \arrow{d}{(e^{2 \pi i x}, e^{2 \pi i y})}\\
Tor^2 \arrow{r}[blue]{T}
& Tor^2
\end{tikzcd}
\end{center}

Taking the parametric equation of the torus $f:\mathbb{R}^2 \to \mathbb{R}^3$:

\begin{equation}
f(\theta,\phi)=((R+r cos(\theta))cos (\phi),(R+r cos (\theta))sin(\phi),r sin(\theta))
\end{equation}

Where $R$ is the distance from the origin to the center of the rotation axis and $r$ is the radius of the circular section of the torus, $R>r$ and $\theta , \phi \in [0,2 \pi)$.

We build a dynamical system on it with initial conditions $(x_0,y_0)$ moving on a straight line (because this is the way in which customers move in aisles in stores) in the direction of $\lambda f_ \theta + \mu f_ \phi$, with $t \in \mathbb{R}$ 

To calculate this, we substituted the parametrized line $\theta = x_0+ \lambda t$ and $\phi=y_0 +\mu t$ in equation 27 getting the following system with respect to $t$; we did this to get a time stamp in the system:

\begin{equation}
x(t)= ((R+r cos(x_0+\lambda t))cos(y_0 + \mu t),(R+rcos(x_0+\lambda t))sin(y_0 + \mu t),rsin(x_0+\lambda t))
\end{equation}

Taking the rotation number on the torus is as follows:

If $T:S^1 \to S^1$ is a measure preserving homeomorphism and $f$ the function that represents it, then the rotation number is:

\begin{equation}
\alpha = \alpha (T) = \lim\limits_{ n \rightarrow \infty} \frac{f^n(x)}{n}(mod 1), x \in \mathbb{R}
\end{equation}

where $f^n$ is the iteration $n$ of the system.

If the rotation number is rational then we have recurrent orbits on the system and if it is irrational the orbits will be dense on the torus, that is why we will take transformations with rational rotation numbers, which are the ones that map trajectories of customers in stores.

The calculation of the probability of collision will be the same as the number of intersections of trajectories on the torus at a given time.

This assumes that the customer is moving at a constant rate given by $f$.

This with other cases, like moving backwards, change in speed, etc. will be addressed with the simulation engine.

\subsection{Defining trajectories}
\label{sec:trajectories}

The best way to define trajectories of customers in a store is the use of an expectation maximization algorithm. This is aligned with  \cite[p.~1096]{kaluza2010}  %[Kaluza, 2010 or 2015, 1096] 
as they divide the ships movements according to ship types. In this case, we divide the customers according to their type. The problem here is to define the type as it is hidden in the basket contents. 

Out of basket information, a matrix of customers should be built with the products associated to them. An important assumption is that a customer = basket. In this way, the association of all the products in a store is directly mapped to specific customers, given their baskets.

\begin{center}
\begin{math}
\begin{bmatrix}
P/C & C_{1} & C_{2} & C_{3} & \hdots & C_{n}\\
P_{1} & 1 & 0 & 0 & \hdots & 1 \\
P_{2} & 0 & 1 & 0 & \hdots & 1 \\
P_{3} & 0 & 0 & 1 & \hdots & 0 \\
\vdots & 1 & 1 & 0 & \hdots & 1 \\
P_{m} & 0 & 0 & 1 & \hdots & 0 \\
\end{bmatrix}  
\end{math}
\end{center}

Where $C_i$ is the $i^{th}$ customer for any $i \in \mathbb{N}$ and $P_i$ is the $i^{th}$ product for any $i \in \mathbb{N}$ 

In this way customers are vectorized with respect to the products in their baskets. Each customer is represented by the column vector of the matrix.

The next step was to assess some similarity in order to cluster the most typical shopping trips in a store which will be mapped to trajectories or orbits in the dynamical system of the store.

For this we used the cosine similarity defined by the inner product of vectors:

\begin{equation}
cos(\theta) = \frac{u_1v_1+u_2v_2+...+u_mv_m}{\| \overline{u} \|\| \overline{v} \|}
\end{equation}

where $\overline{v}$ and $\overline{u}$ are any two m-dimensional vectors with coordinates $\overline{v}=(v_1,v_2,v_3,...,v_m)$, $\overline{u}=(u_1,u_2,u_3,...,u_m)$ and $\| \overline{v} \| = \sqrt{v_1^2+v_2^2+v_3^2+...+v_m^2}$

The next step is to apply an Expectation Maximization (EM) algorithm to the matrix of cosine scores. This is to create clusters to identify the $n$ most common baskets and their respective journeys. Using customer data from loyalty schemes would provide detailed basket groupings, however an alternate solution like EM (an unsupervised learning algorithm) will provide adequately accurate groupings. As a result, customer privacy will be kept intact whilst providing sufficiently accurate data for the simulation. 

For implementation purposes, the \emph{Scikit-Learn} library is being considered. Specific details can be found at  https://scikit-learn.org/stable/modules/mixture.html

\subsection{Minimum sample size}

One of the main questions in the simulation process is to calculate the minimum amount of simulations that are needed to achieve the desired results for the model to be used confidently.

In order to determine the minimum sample size we refer to the classic theory of statistics, where the objective is to calculate this parameter to estimate the population mean. 

As the theory dictates, the objective of this calculation is to get intervals as narrow as possible aiming for the highest reliability. 

For this, we need several factors: i) the confidence interval $1- \alpha$, necessary to calculate the  ii) reliability coefficient $z$ which is $z_\alpha = \frac{x- \mu}{\sigma}$ for a normal distribution for a given significance level $\alpha$ and iii) the standard error $SE= \frac{\sigma}{\sqrt{n}}$, where $\sigma$ is the standard deviation and $n$ is the sample size.

In this case, $\sigma$ is a fixed value, given by the data itself, therefore the only value to reduce the standard error is to increase the value of $n$.

The length $l$ of the confidence interval is:

\begin{equation} \label{eq:conf_interval}
l = z SE = z \frac{\sigma}{\sqrt{n}}
\end{equation}

As the population treated in this case is a large population (the entire UK population), we will ignore the population correction $\sqrt{\frac{N-n}{N-1}}$ where $N$ is the population size. Hence solving equation \ref{eq:conf_interval} for $n$:

\begin{equation}
n = \frac{z^2 \sigma^2}{l^2}
\end{equation}

If we consider the special case of a smaller population to simulate, for example a local population outside a densely populated urban area, we would need to multiply equation \ref{eq:conf_interval} by $\sqrt{\frac{N-n}{N-1}}$, obtaining the minimum sample size $n$:

\begin{equation}
n = \frac{N z^2 \sigma^2}{l^2 (N-1) + z^2 \sigma^2}
\end{equation}

In general $\sigma^2$ is an unknown, but it will be assumed that the population is approximately normally distributed and consider $\sigma = R/6$ as in the 6-sigma theory, where it is consider that the entire range $R$ of a normally distributed population is $R = 6 \sigma$.

As an example the ideal situation would be to measure the amount of footfall in a supermarket in a day with an error of $\pm$ 100 people. We assume that the standard deviation $\sigma$ for the number of people in a supermarket is 200 people, and the confidence level $\alpha$ is 5 percent, then we obtain the following sample size:

\begin{equation}
n = \frac{z^2 \sigma^2}{l^2} = \frac{1.96^2 200^2}{50^2} =61.47
\end{equation}

So 62 simulations would be required to achieve statistical significance given the prior assumptions of $\sigma$ and $\alpha$.

\printbibliography
% \begin{thebibliography}{9}

% \end{thebibliography}

\section{Technical Appendix}
In this section we give an overview of the simulation engine that we have developed to run at scale across multiple store geometries and basket distributions. Results and conclusions of this work will be presented elsewhere.
\subsection{Simulation and randomness}

There are a number of factors that are very difficult to model with mathematical theory due to the randomness of their occurrences. These cases are better estimated with a simulation engine. 

The random cases that the simulation will contemplate are on the following list. The decision of including them or not, depends on the significance and impact on the results compared to the complexity of including them in the model.

\begin{center}
\begin{tabular}{ |l|c| } 
 \hline
Feature & Included?\\
1. Customer able to move at different speeds & Y/N \\ 
2. Customer may get slower as items are added to basket & Y/N \\ 
3. Customer speed gets a penalty when adding a heavy item (stacks) & Y/N\\
4. Customer will pick up item and later put it aside & Y/N \\
5. Customer will get distracted by certain items & Y/N \\
6. Customer will have baggage & Y/N \\
7. Customer will move at different speeds in different parts of the store & Y/N \\
8. Customer will not add item to basket but will later come back to it & Y/N \\
9. Customer will wander, occasionally picking up and putting items back & Y/N \\
10. Customer will avoid certain aisles in the store & Y/N \\
11. Customer will not use the basket, but the trolley in the store & Y/N \\
12. Customer will use multiple baskets when no trolleys are available & Y/N \\
13. Customer will return the trolley at the end of the shop & Y/N \\
14. Customer will leave the trolley near the car &  Y/N \\
15. Customer will return to the store whilst in the Queue & Y/N\\
16. Customer will return to the store whilst in Checkout &  Y/N \\
17. Customer will periodically return to a Bay/Aisle because it is inaccessible & Y/N \\
18. Customer will return the next day if item is out of stock & Y/N \\
19. Customer will add cold items last & Y/N \\
20. Customer will pick up multiple copies of the same item before choosing one & Y/N \\
21. Child Customers will touch multiple products that the Parent Customer will put back & Y/N \\
22. Child Customers will sit on the floor and will later be carried by the Parent & Y/N \\
23. Customer will shoplift & Y/N \\
24. Customer will pay by cash & Y/N \\
25. Customer will pay by card & Y/N \\
26. Customer will pay by contactless & Y/N \\
27. Customer will violate Physical Social Distancing & Y/N \\
28. Customer will deliberately attempt to infect other Customers & Y/N \\
29. Parent Customer will shout / talk loudly to the Child Customers when they misbehave & Y/N \\
30. Customer will need a parking ticket for the car park & Y/N \\
31. Customer will abandon shopping midway, leaving basket where it is/entrance & Y/N \\
32. Customer will abandon shopping midway, returning items to bays & Y/N \\
33. Customer will wear gloves and / or masks & Y/N \\
34. Customer will take off / put on gloves and / or masks midway & Y/N \\
35. Customer will open a pack, consume its contents for the remainder of the time in store  & Y/N \\
36. Customer will open a pack from a bay, consume a portion of its contents and put it back & Y/N \\
37. Customer will meet up with other Customers in the store and shop together & Y/N \\
 \hline
\end{tabular}
\end{center}

\subsection{Metrics collected}

The metrics to be collected from the simulation engine are divided in four sections, counters, timers, gauges and sets:

1. Counters. \par

a) One important metric, which actually helps to estimate the rotation number of the dynamics in the store is the number of times a customer goes to a location. This will also let us know about the ``hot spots" or most frequent places shoppers go in a store.

b) The number of Customers at a time in a store. This metric will help to estimate the maximum number of customers in a store so the system does not gridlock and also maximizing the number of customers served in a given time. 

c) The number of Customers at checkouts at a time

d) The number of Customers shopping at a time

e) The number of Customers idle at a time

f) The number of Customers waiting at a time (queuing)

g) Half life calculations can perfectly indicate when a store will need replenishment or other type of maintenance e.g. collecting and disinfecting trolleys. The calculation of the time when a half empty store (product-wise) happens is vital to schedule replenishment shifts. This will help to shape policy on when staff has to go to the shop floor and replenish or fulfill other duties and interact with customers.

Obviously, this will increase the traffic and congestion, but only until the replenishment is done. 

h) Another important measure is the half empty store (customer-wise) as will also be a good indicator for synchronizing staff duties and minimize disruption.

All these measures might be able to calculate ``close calls" or  ``near misses" on collisions which will indicate dangerous situations hence helping shape policies for shopping in any store. This will happen when a customer does not respect the 2-meter distancing rule.

2. Timers. \par
a) Time at checkout

b) Time at shopping

c) Idle time 

d) Waiting time

e) Total time in the store

3. Post simulation metrics. \par

a) Number of customers infected after a simulation

b) Number of customers at risk after a simulation

c) Number of customers with antibody after a simulation

\subsection{Unity 3D: Simulating A Supermarket}

\subsubsection{What is Unity?}

Unity is a platform to design games developed by ``Unity Technologies". This platform or game engine, is used to develop video games that can be run in different environments: web plugins, desktop, consoles and mobile devices.

The main purpose of using Unity to design the simulation is that it allows to build applications where the ``visible pieces” of the simulation can be put together with a graphical preview using a controlled ``play it” function.

\begin{figure}
\centering
\includegraphics[width=0.9\textwidth]{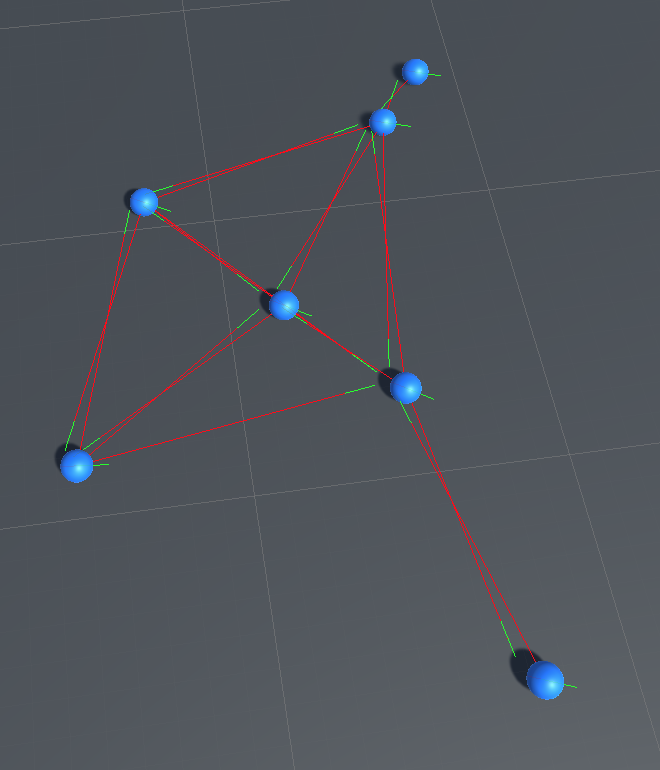}
\caption{Unity (DOTS) implemented using Pure ECS. Spheres are observing neighbouring spheres (red rays), calculating the distance between them each frame. Green rays here indicate surface normals for distance calculation and orientation.}
 \label{fig:unity-dots-physics-1}
\end{figure}

In addition, Unity allowed for a quick and simple test to be conducted on the performance and allowed running multiple simulations in parallel. See Figure \ref{fig:oo-sim-nine} for an example.

The code was written using the existing system of \code{Monobehaviours} \& \code{GameObjects} which is also referred to as the Object-Oriented style of development in Unity.

Finally, Unity can import 3D models and is able to simulate physics in relation to Newtonian Mechanics, allowing it to model and simulate colliders, rigid bodies and kinematics.

\subsubsection{The Environment}

Each environment consists of Agents, a Spawn point, and a store, where an Agent represents a person in a store. The Spawn and Despawn points, are locations within the environment where the Agents are added to the simulation and removed at the end. Each store, with all the elements of a common supermarket are included based on the floor plans. This includes: aisles, tills, exits, entrances etc.  All these elements were converted to a \code{prefab}. 

The first implementation consisted of using the inbuilt \code{Animator} to model and execute the Finite State Machine that controls each customer, where a Finite State Machine is a computation model used to simulate sequential logic.

\subsubsection{The Simulation: How It Works}

Note: to standardize terms, we will refer to agents, people and customers as the same. 

Customers are spawned from the Spawn point and upon spawning, are tasked with visiting five bays (initially chosen at random, later to be guided by historical transaction data as described in Section  \ref{sec:trajectories}) .

Once these five bays had been visited, the agent is asked to queue before heading to the checkout tills. A checkout till was chosen at random, no specific queuing behaviour was included at this early stage. After completing the checkout stage, agents head to the Despawn point where they are removed from the environment.

\begin{figure}
    \centering
    \includegraphics[width=0.9\textwidth]{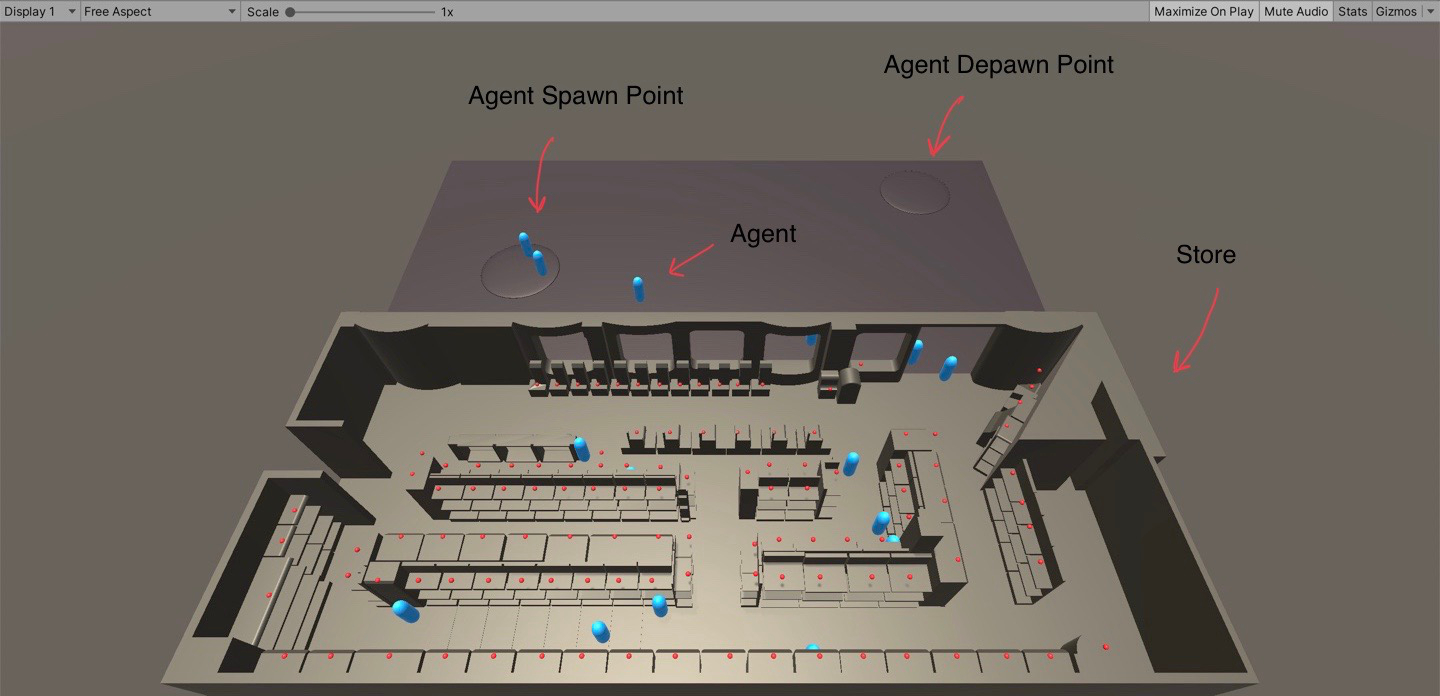}
    \caption{Store in Unity. Agent Spawn \& Despawn Points marked. Agent is modelled in blue. }
    \label{fig:oo-sim-store}
\end{figure}

All tests had identical starting hyper-parameters.

The Customer Spawn point was set to spawn 50 agents (in total) at a rate of one customer every four seconds.

\subsubsection{Technical Notes and Limitations}

\begin{figure}
    \centering
    % TODO: Mark the spawn and despawn points in the image
    \includegraphics[width=0.9\textwidth]{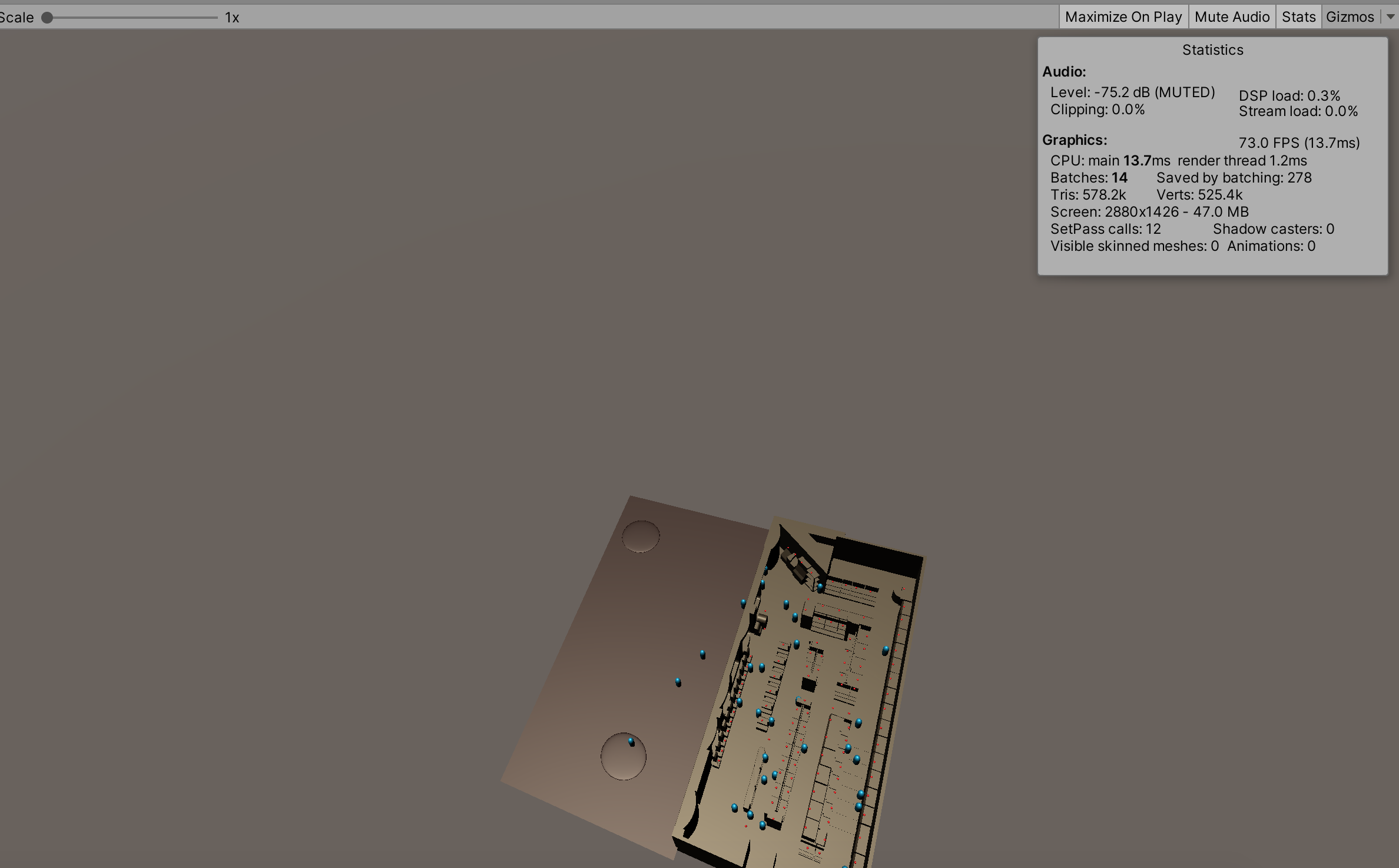}
    \caption{Single simulation. Frame rate hovers around the 60 FPS mark. 
The statistics shown in the top right hand corner are specific to Unity's Game environment, they do not cover the simulation itself, like the number of infected agents, but this can be calculated using the techniques described in sections above.}
    \label{fig:oo-sim-single}
\end{figure}

When running a single simulation, the Frames per Second (FPS) remained at a healthy level, hovering around the 60 FPS mark. This is done to respect the physics of the environment. Different speeds would affect the accuracy of the physics  engine and lead to a far-from-reality environment. See Figure \ref{fig:oo-sim-single} for a screenshot showing this.

\begin{figure}
    \centering
    \includegraphics[width=0.9\textwidth]{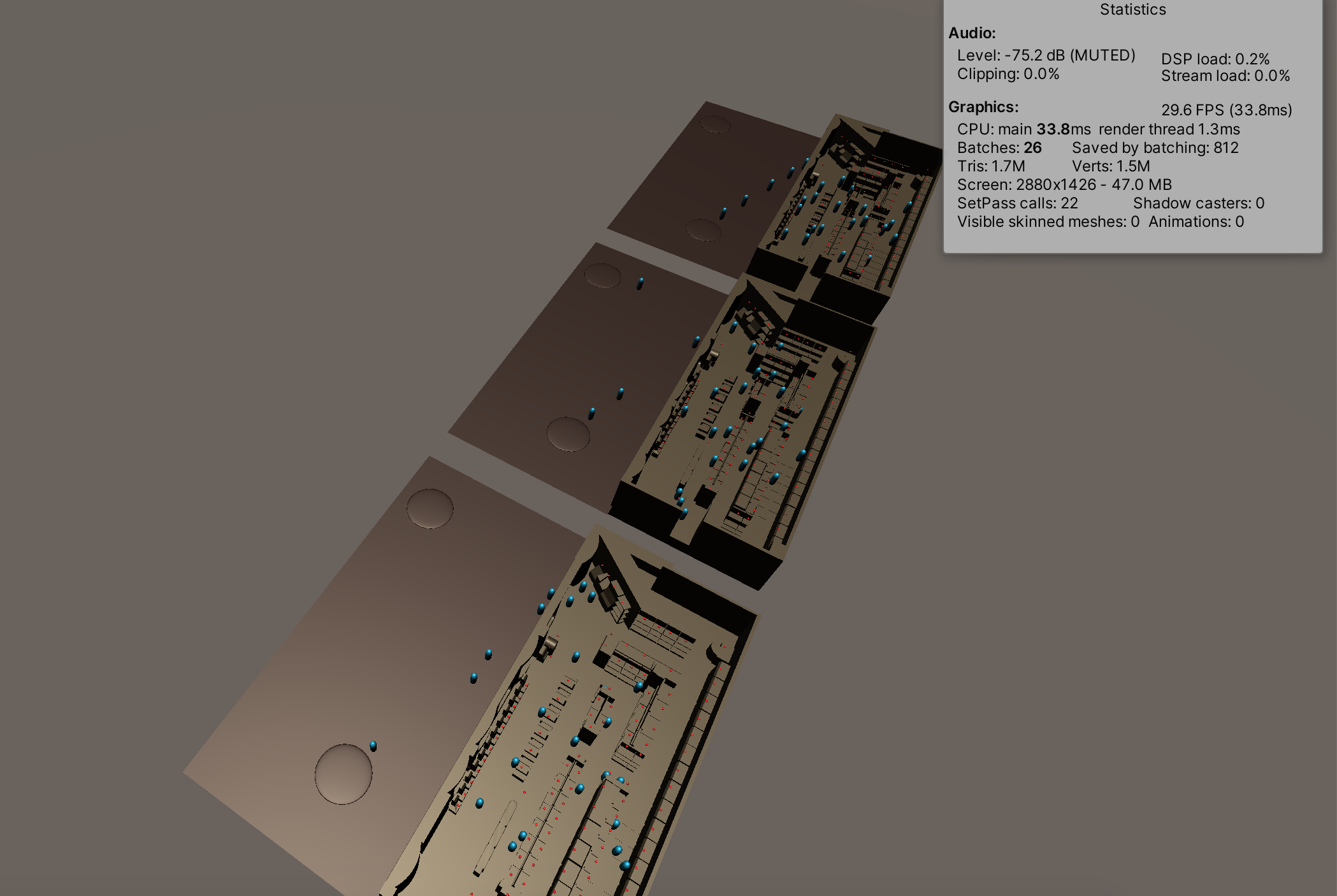}
    \caption{Three parallel simulations. FPS drops to 28 with noticeable jitter in the movement of the agents, particularly during collisions throwing the accuracy of the collision data into question. Note that this jitter is different from that seen when two \code{RigidBody GameObjects} with \code{Colliders} collide in a narrow passage. }
    \label{fig:oo-sim-triple}
\end{figure}

When adding multiple simulations in parallel, the performance of the engine took a drastic hit. In the case of three parallel simulations, the frame rate dropped to 28 FPS (Figure \ref{fig:oo-sim-triple}). 

This is an acceptable frame rate for the run if it didn't drop occasionally to single digits during moments of high collision activity.

\begin{figure}
    \centering
    \includegraphics[width=0.9\textwidth]{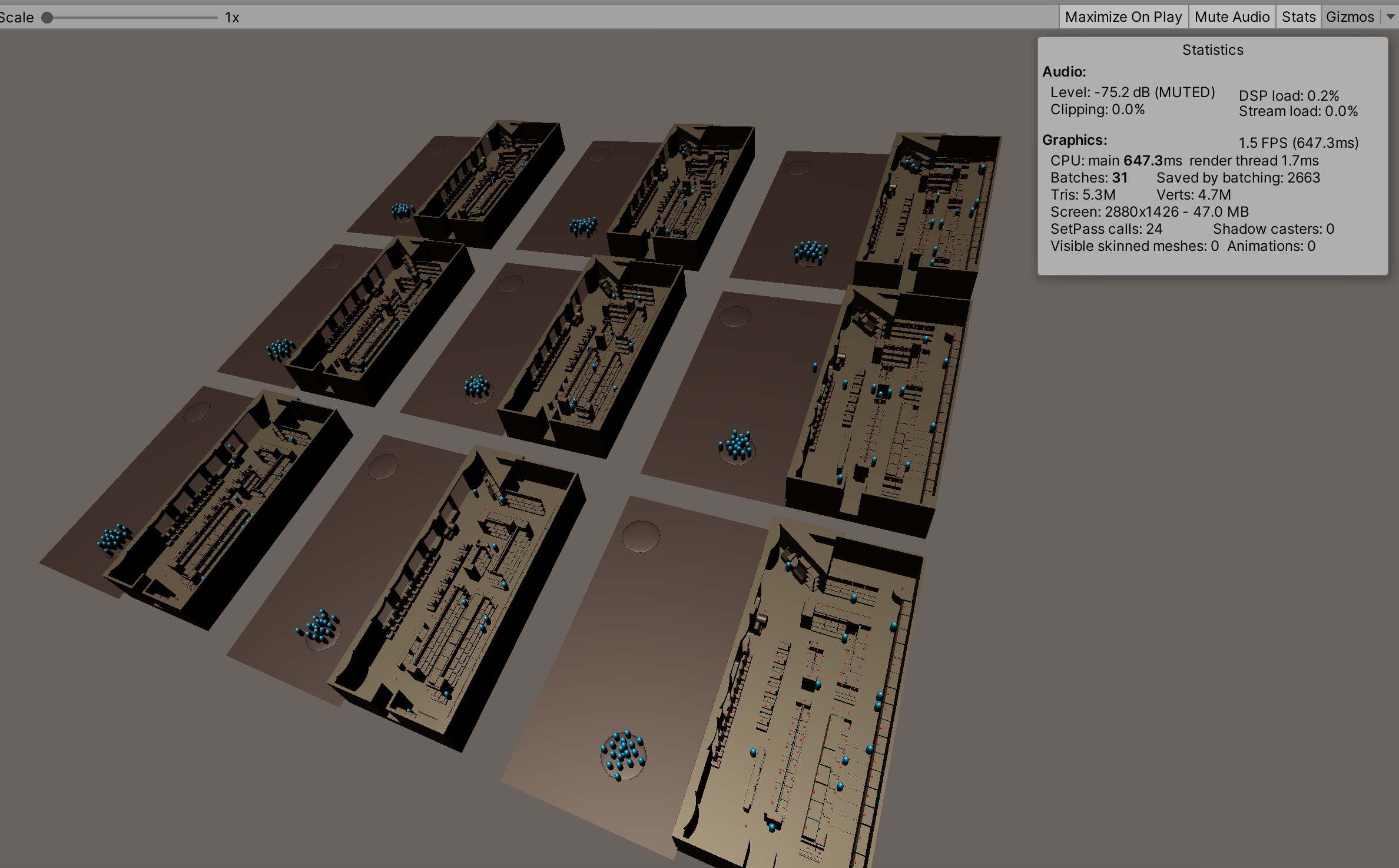}
    \caption{Nine parallel simulations. FPS is between 1 and 2 making the data completely unusable. Several spawned agents can be seen waiting at the spawn point for instructions.}
    \label{fig:oo-sim-nine}
\end{figure}

When running nine parallel simulations the rate dropped to 1.5 FPS with each frame taking (on average) between 600ms and 800ms (Figure \ref{fig:oo-sim-nine}). This makes it unusable for analysis purposes.

When deploying in headless mode (also known as a `Server build') the performance is expected to improve although it is unlikely to be by the one or more orders of magnitude that is needed for efficient and reliable experiment data gathering. 

Moreover, with this jerky frame rate, the accuracy of the collision data comes into question as agents can `skip' or `glitch' past each other instead of colliding.

An easy solution to this might be to drop the accuracy of the PhysX engine solver powering the rigidbody mechanics in the simulation, but that would also compromise the integrity of the collision data, making it unviable.

\subsection{Tackling limitations: Scale Up and Production}

In order to generate the synthetic data which at an aggregate represents real-world
behaviour, we need to run millions of simulations. While Unity provides a high-
performance solution to aid in simulation and some capabilities to run it
simultaneously on a single machine, we need a solution to run hundreds and
thousands of simulations in parallel in a timely and cost-effective way. Additionally,
we need the environment to run experiments with different hyper-parameters and
enable traceability of simulation results all the way back to the parameters which
were used to generate it.

The general goals of the solution are to:

- Run simulations at massive scale. \par
- Execute simulations in a cost effective and timely way \par
- Store simulation data into a performant backend \par
- Support repeatable \& traceable process for running experiments \par
- Monitor simulations and gain real-time visibility to track blockers or parameter issues \par

\subsubsection{Running Simulations At Massive Scale} 

To scale the simulations the project leverages a cloud based solution.

Given the large-scale compute requirements, leveraging a public cloud was a pragmatic decision. Google Cloud Platform (GCP) has compelling technical solutions to address the simulation needs and was a prime candidate for the solution. Note that it is possible to replicate the solution to other cloud environments.

The key component to address the scaling requirement is Kubernetes (commonly referred to as K8 in the community). Kubernetes is an open-source container-orchestration system for automating application deployment, scaling, and management. It was originally designed by Google and is now open sourced and maintained by the Cloud Native Computing Foundation (CNCF). Kubernetes can be configured to deliver a “serverless” batch model providing an on-demand burst capacity at lower cost i.e. we don’t have to pay for idle capacity and allows us to leverage the massive scale of the cloud to run simulations in parallel. Google Cloud Platform provide a managed Kubernetes service Google Kubernetes Engine (GKE).

Using GKE enabled us to quickly setup the environment and not have to worry about administering/setting up the cluster from scratch.

\subsubsection{Stream Simulation Data Into A Performant Storage}

While the Kubernetes cluster in the cloud would allow us to massively scale the compute for simulations, we also require a scalable storage solution for the simulation data. Given a single simulation can span days the desire was for the data to be streamed in real-time as opposed to an end-of-simulation batch update. The other benefit of real-time streaming is that it provides a steady throughput to backend storages vs a spikey end of run batch data. Google cloud provides several storage options to handle large data including Cloud Storage, BigTable, Spanner and BigQuery. Given the structure of the data and considering the pattern of how the simulation data would be used subsequent to data ingest, Google BigQuery was identified as the best fit. BigQuery is a serverless, highly scalable, and cost-effective cloud data warehouse solution. Additionally, BigQuery natively supports streaming data to simplify the ingest process.

\subsubsection{Repeatable \& traceable process for running experiments}

We can have several experiments running in parallel and this would produce a high volume of simulation data, it is imperative for us to have a solution that would provide traceability and lineage of the hyper-parameters and the corresponding version of the Unity code that was used to generate the data. The following design choices were made:

1)	Docker: Docker enables us to package the Unity simulation code application along with all of the dependencies into a self-contained image. Container based deployment is at the heart of Kubernetes and made Docker a simple choice. Google Cloud provides Google Container Repository (GCR) as a central Docker image hub. Code and dependencies for the Unity solution once pushed into the code repository can trigger a continuous delivery pipeline that would create a Docker image and push it into GCR. The image with its label would be pulled by the Kubernetes job to run the simulation.
 
2)	Job Scheduler:
The environment needs to support both exploratory runs of the simulation to tune hyper parameters and also for the full run of simulations with the target hyper-parameters. Kubernetes has a native template to run batch jobs where we can specify the number of parallel runs however with the in-built solution, we cannot run simulations with different parameters. A custom job scheduler solution is required to take the experiment parameters (from the experiment manifest) and transform it to one or more jobs based on the combinations of different parameters. The scheduling solution also stores the parameter and job details to the underlying back end storage so that it can be used for traceability and other queries. The process allows for a clean separation of hyper-parameters which are specified in the experiment manifest vs. mapping to Kubernetes technical parameters which are handled by the job scheduler. Note: Batch on Google Kubernetes Engine (GKE) allows for array parameters and would be a good candidate however, at this point in time batch on GKE is in early beta and was not considered

\begin{center}
\includegraphics[width=80mm]{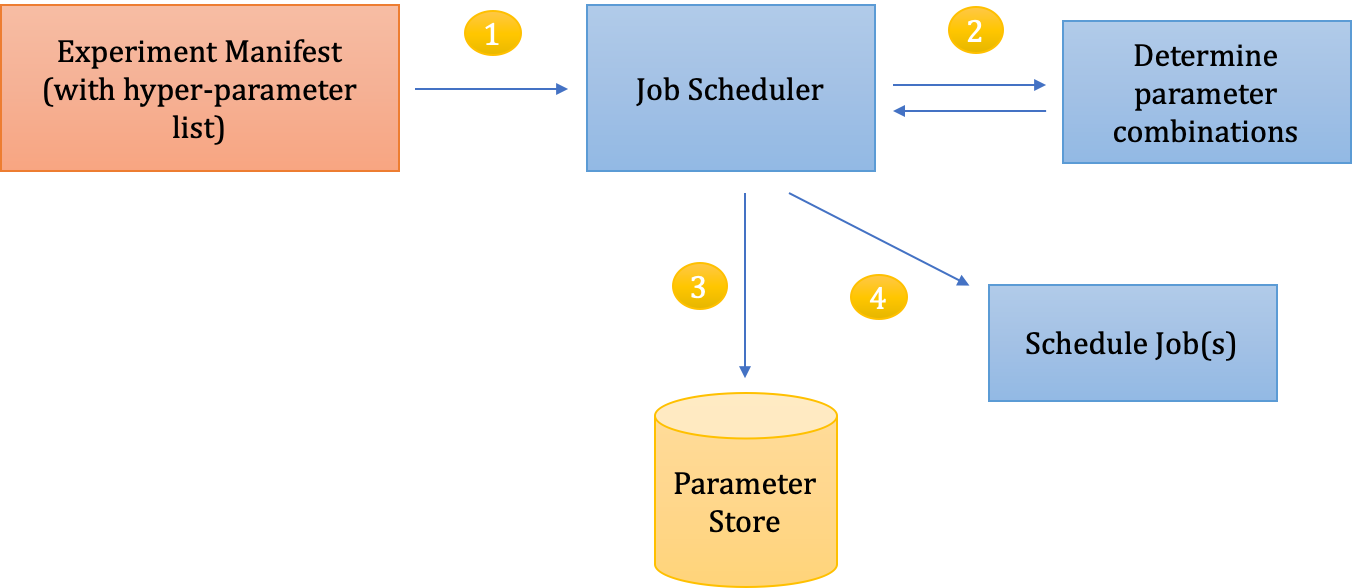} \par
Job scheduling for an experiment.
\end{center}

3)	Parameter tracking in simulations:
Since the simulations are run in a distributed cluster with support for asynchronous parallel executions, we need a mechanism to track simulations independent of each other and track the parameters used to trigger the simulations. The Job scheduler would provide the experiment id and the job id along with hyper-parameters to the Docker container. The Unity code would access the values via Environment variables or command line parameter and generate a simulation id e.g. use a Universally Unique Identifier (UUID). The simulation data would be stored along with the combination of keys in the underlying database tables.

\begin{center}
\includegraphics[width=80mm]{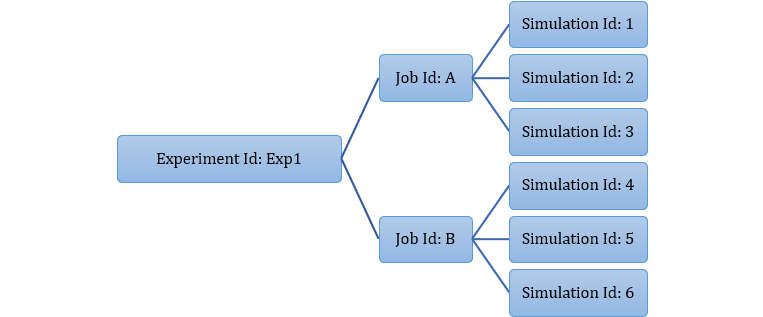} \par
Logical hierarchy for capturing simulation data with sample ids.
\end{center}

\subsubsection{Monitor and track simulations in real-time}

Generating simulations require several hyper-parameters to be provided to the container. There is a potential for the combination of some parameters to disrupt the simulation or cause the simulations to generate incorrect data. We need a mechanism to monitor the simulation across the parallel executions and give us real-time visibility into simulations that quickly detect if any of the simulations are running incorrectly. In addition to the simulation data there are other useful metrics and events data that is generated which may be useful at later stage. The sheer volume and velocity of manging this data presents a different challenge to the challenge of manging the primary simulation data. To address the capture and monitoring of the data a separate real-time time-series database was required. InfluxDB was identified as a suitable candidate and Grafana was identified as a good choice for visualising the real-time data. Google cloud natively provides logging and metrics capture for technical metrics such as CPU/memory etc hence no additional solution was used to capture/monitor technical metrics.

\subsubsection{Overall Solution}

\begin{center}
\includegraphics[width=80mm]{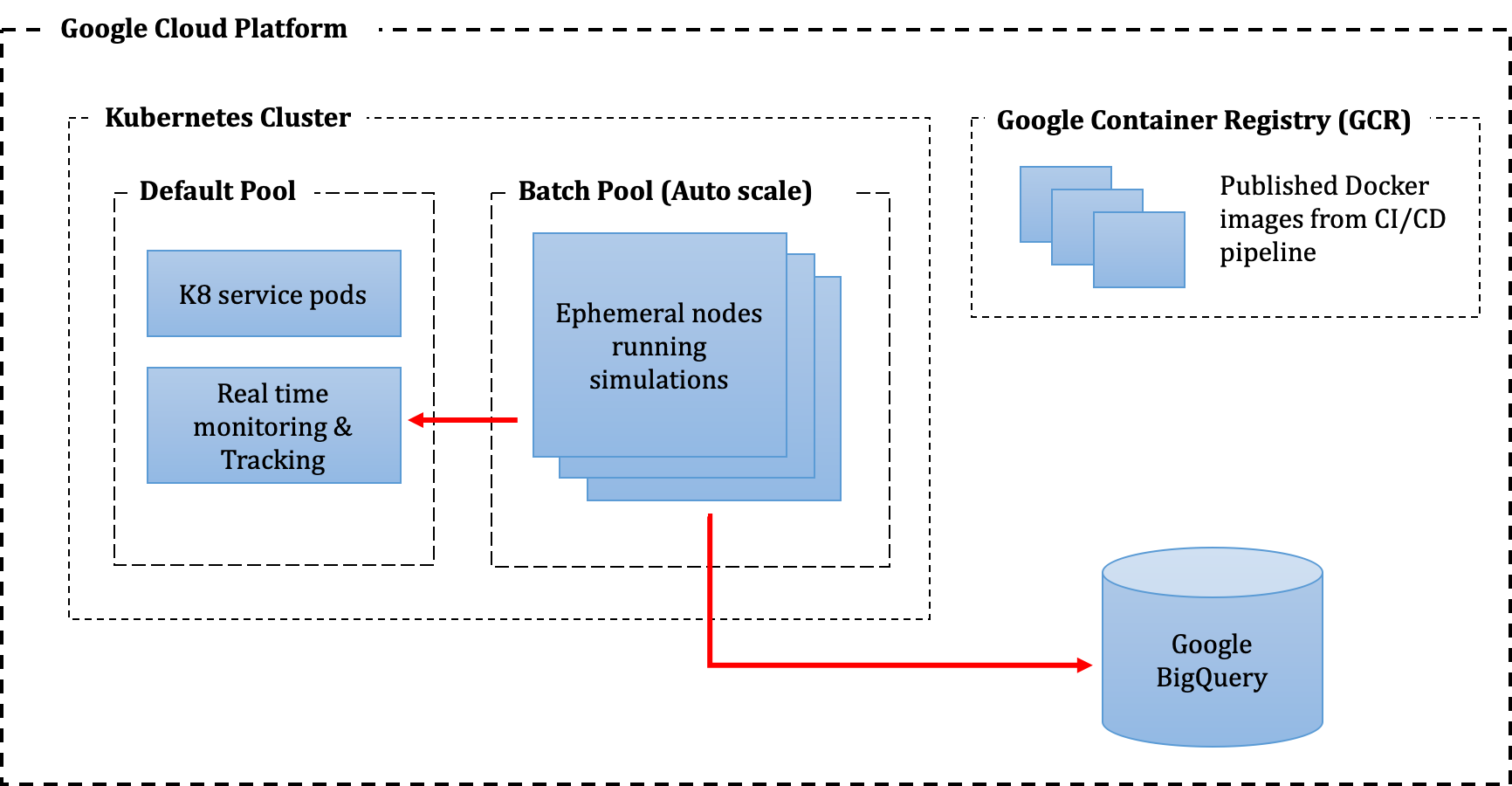} \par
High level solution
\end{center}

Key design decisions:

-	Default Pool: The default pool would have a fixed number of server nodes. The default pool would be used to manage the core Kubernetes pods (i.e. kube-system namespace) as well as some of the monitoring and tracking related pods. Apart from the InfluxDB pod none of the other pods required significant resources hence a lower spec infrastructure was used for the default pool. Helm charts are used to produce the templates for deploying the InfluxDB and Grafana services.

\begin{center}
\includegraphics[width=80mm]{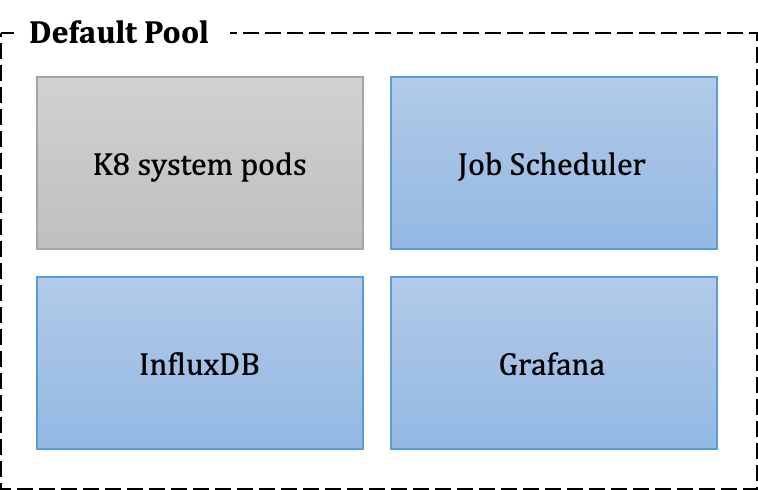} \par
Conceptual view of the Kubernetes cluster default pool
\end{center}

-	Batch pool with autoscaling: A separate Kubernetes pool, “Batch pool” was created with auto-scaling turned on and minimum node size set to zero. The nodes in this pool would effectively provide a pool of ephemeral nodes to service the simulation jobs. When the nodes in the pool are idle for a while the Kubernetes autoscaling will automatically de-provision the nodes in the background. The nodes spec for the pool was for set high compute. The pool configuration also specified the taints on the nodes in the pool to prevent other pods such as the Kubernetes kube-system namespace pods, InfluxDB and Grafana pods from being run in the pool. The Job scheduler would be used to specify the tolerance on the job template to include the taint key for the batch pool to ensure that that the nodes in the batch pool are used only for the simulation jobs.

\begin{center}
\includegraphics[width=80mm]{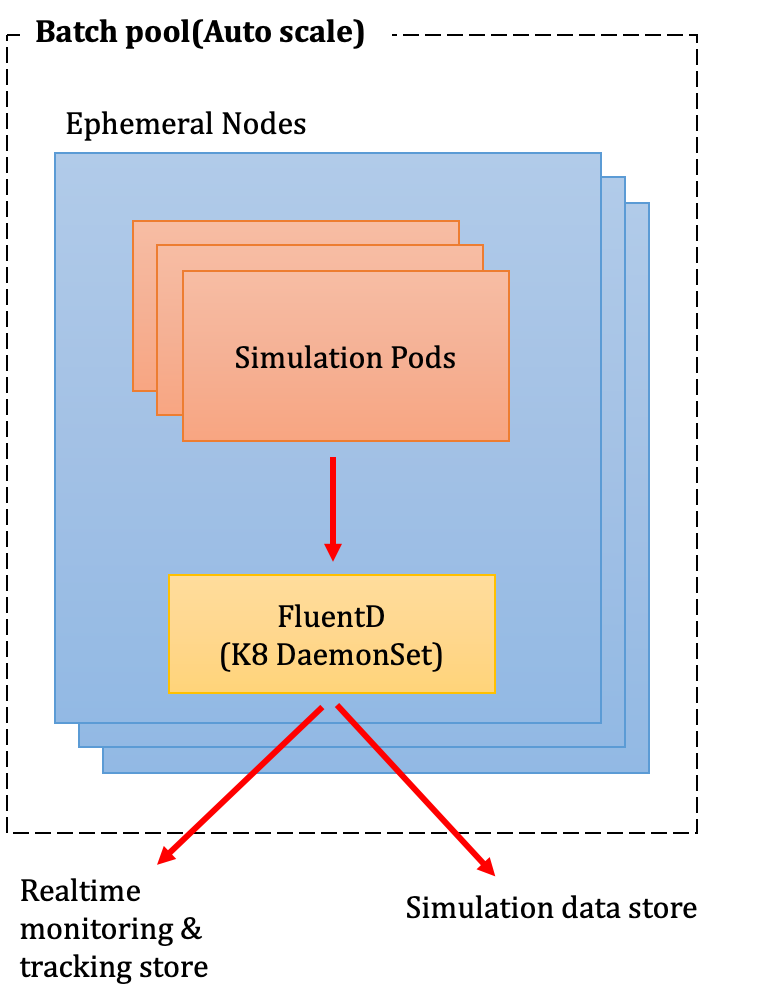} \par
Conceptual view of the batch pool
\end{center}

-	Data capture and ingest: The project leverages FluentD. FluentD is an open source data collector and is part of Cloud Native Computing Foundation (CNCF) and comes integrated with Google Kubernetes Engine. The simulation code is integrated with FluentD library to write logs which are picked up by the FluentD pods which are configured to run as a Kubernetes DaemonSet. FluentD has plugins for Google BigQuery and InfluxDB. FluentD provided a config driven approach to route the data and handle different load conditions. There are other options in Google cloud like using Pub/Sub and Cloud Functions however FluentD was selected as it was a much simpler option (fewer moving parts) and reduced the cost of the overall solution.

\subsection{Further development}

This paper serves as the foundation for the practical work with real data. As a next step we need to run the simulations with actual data that should be provided by supermarkets or other places. 

Another important next step is the production at scale of simulations. This might also deserve a separate study in itself. At the time of this performance test, Unity Technologies is actively developing a more performant solution known as the Data Oriented Technology Stack or DOTS. Using this approach we have seen significant improvements that a) allow for over a thousand agents to be active in a single simulation, b) complex distance queries to be performed by the Unity Physics engine that were previously difficult to parallelise. See Figure \ref{fig:unity-dots-physics-1} for an example of the system in action. Whilst this approach will require more detail in implementation, it comes with the ability to perform far more detailed simulations where we have greater control of the number and behaviour of the agents.

As a final remark, we realized that this work could not have been achieved without the proper ensemble of a multi-disciplinary team that included mathematicians and computer scientists and the intelligent iterations of all the individual efforts put together as a team.

\subsection{Addendum: (Abandoned) Infection Spread in Unity}

Towards the start of the analysis, there was an attempt to model the spread of infections in the store.
Each agent was given an \code{infected} and \code{infectious} boolean attribute.
Along with this, the infectious agent was given an infection probability, which roughly equated to the chance that the agent could pass on the infection to a vulnerable, uninfected agent, and an infection radius which represented the agent's ability to spread the infection. 
Once a healthy agent was infected, its \code{infected} status was set to \code{true} but as the infection cannot be passed on instantaneously the \code{infectious} property was left as \code{false}.

\begin{figure}
    \centering
    \includegraphics[width=0.9\textwidth]{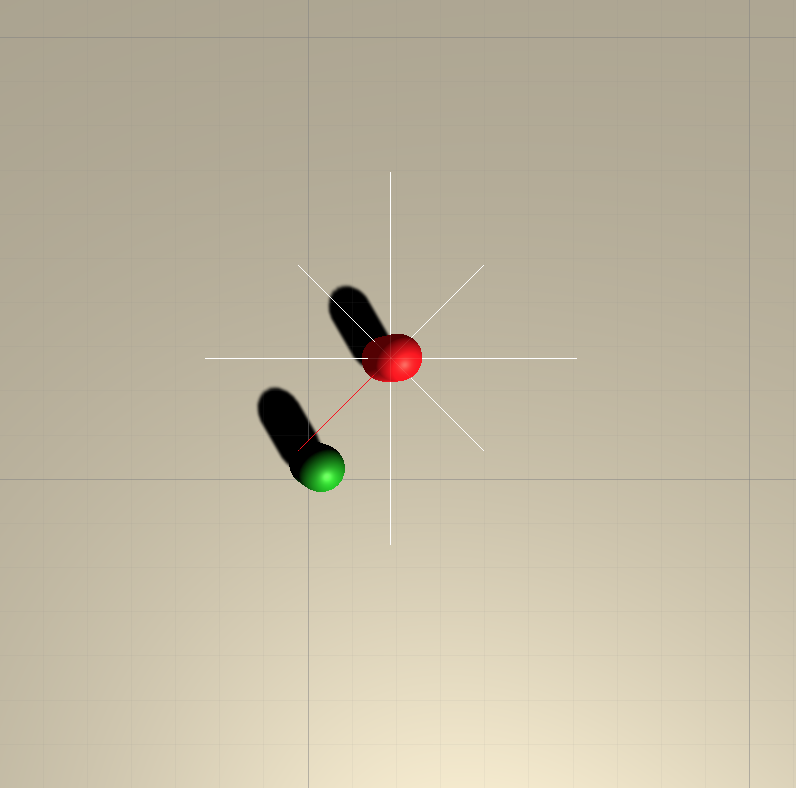}
    \caption{An early, but flawed attempt to model the spread of the infection within Unity using \code{RayCasting}. Here the infected agent (in red) is casting eight rays around itself in 45$^{\circ}$ increments. The length of the ray is set to the agent's \code{infection radius}. Different agents have different infection radii depending on how infectious they are. If another agent comes within the length of this ray then it (the other agent) is considered `exposed' to the infection and has a chance of catching the infection from the infected agent that is equal to the infection spread probability of the infected agent. Each agent has a \code{SphereCollider} which surrounds the agent with a radius equal to the social distancing radius. This allowed for simulations to check how different social distancing radii affected the spread of the infection based on infection radii and infection probability. Ultimately, \textbf{this attempt to model the infection spread was abandoned} as it is not known if the virus spreads in this manner and it could lead to faulty experiment data.  }
    \label{fig:oo-infection-spread}
\end{figure}

Several methods for implementing this `spread' were considered.
Using the inbuilt \code{Collider} mechanism to \code{Trigger} when two agents came within the infection radius would work but it could prove to be difficult (but not impossible) to model different social distancing radii with different agent infection radii. 
A \code{RayCasting} approach was taken instead (see Figure \ref{fig:oo-infection-spread} for an example).

Eight rays were cast outwards from the agent in forty-five degree spreads.
The length of each ray was set to the infection radius. 
If another agent came within the length of the ray then that agent was considered `exposed' to the infection. 
However, as noted this did not mean the agent became infected immediately as there was a probability of the exposed agent catching the infection.

Ultimately, \textbf{this attempt to model the infection spread was abandoned} as it is not known if the virus spreads in this manner and it could lead to faulty experiment data. 
For instance, if the spread of the virus is directed in a specific direction based on the air currents in the store, this primitive ray casting method would not be able to accurately model the spread. 
A ray can only travel in a straight line and it is unlikely the aerosol transmission of the virus would follow such a restriction. 
Following a different approach of recording all the collisions and their durations (as detailed in Section \ref{sec:probability}) is the preferable option as it allows for analysis as more data about the aerosol dispersion of the virus is researched.

\end{document}